\documentstyle[aps,amssymb]{revtex}

\begin{document}
\title{Level Correlations And Persistent Currents In Mesoscopic Metals}
\author{S. Sitotaw and R.A. Serota\thanks{%
serota@physics.uc.edu}}
\address{Department of Physics\\
University of Cincinnati\\
Cincinnati OH 45221-0011}
\maketitle

\begin{abstract}
We use the exact correlation function of the density of energy levels in the
magnetic field to evaluate persistent currents in mesoscopic metals. We also
analyze the perturbation theory limit of the correlation function vis-a-vis
the perturbation theory limit of the orbital response.
\end{abstract}

\section{Introduction}

Orbital magnetism and persistent currents in mesoscopic metal rings and
grains have received much attention in the past decade. Theoretically,
mesoscopic orbital magnetism is rather well understood for temperature $T\gg
\Delta $, where $\Delta $ is the mean energy level spacing, and for magnetic
fields such that the flux through the sample is\footnote{%
For a narrow ring, the response is $\phi _{0}/2$ periodic so there is no
need to consider larger fluxes.} $\phi \lesssim {\phi }_{0}=hc/e$ - the flux
quantum\cite{Oh Zyuzin Serota} (${\phi }_{0}=2\pi /e$ in units where $\hbar
=c=1$). To evaluate the thermodynamics quantities at these temperatures, it
is sufficient to characterize the energy spectrum in terms of the level
density $\rho $. In particular, the orbital magnetic dependence of the level
density correlation function, 
\begin{equation}
R\left( \omega ,{{\tau }_{H}}^{-1}\right) =\left\langle \delta \rho \left(
\varepsilon _{1},H\right) \delta \rho \left( \varepsilon _{2},H\right)
\right\rangle  \label{R}
\end{equation}
where $\delta \rho \left( \varepsilon ,H\right) =\rho \left( \varepsilon
,H\right) -\left\langle \rho \right\rangle $ and the brackets stand for
disorder averaging, describes both the mean mesoscopic response and the
fluctuations\cite{Oh Zyuzin Serota}.

Furthermore, in this limit it is sufficient to use the perturbative limit of
the correlation function. In the two-Cooperon approximation, it is given by%
\cite{Oh Zyuzin Serota}, 
\begin{equation}
R_{pert}\left( \omega ,{{\tau }_{H}}^{-1}\right) =\frac{s^{2}}{2\beta {\pi }%
^{2}}Re\frac{1}{{\left( -i\omega +\gamma +{{\tau }_{H}}^{-1}\right) }^{2}}
\label{R_pert}
\end{equation}
$\omega =\left( \varepsilon _{2}-\varepsilon _{1}\right) $, $\gamma $ is the
level broadening (which will be neglected below relative to $T$, as it
should for metals), and $s$ and $\beta $ are integers whose values depend on
the symmetry class. For Gaussian Orthogonal Ensemble (GOE) $\beta =1$, while
for Gaussian Symplectic Ensemble (GSE) $\beta =4$ due to the fact that 3 out
of 4 spin channels are eliminated by spin-orbit scattering. Neglecting
Zeeman splitting, $s=2$ in either case due to double spin degeneracy and $%
\left\langle \rho \right\rangle \Delta =s$. ${{\tau }_{H}}^{-1}$ gives the
orbital magnetic dependence. For a single period for a ring with the flux $%
\phi $, it is given by\cite{Oh Zyuzin Serota}

\begin{equation}
{{\tau }_{H}}^{-1}=\left( 2\pi \right) ^{2}E_{c}\left( \frac{2\phi }{{\phi }%
_{0}}\right) ^{2}  \label{tau_H^-1}
\end{equation}
where $E_{c}=\hbar \tau _{diff}^{-1}\gg T$ and $\tau _{diff}=L^{2}/D$ is the
time of diffusion around the ring of circumference $L$ (time of diffusion to
sample boundaries for simply connected geometries, such as disk\cite{Oh
Zyuzin Serota}), $D$ being is the diffusion coefficient.

In the evaluation of the mean response one must remember that the electrons
in a closed system form a canonical ensemble\cite{Kubo}. Ingeniously, Imry%
\cite{Imry} found a simple relationship between the free energy of a
canonical ensemble and the free energy and particle-number fluctuation of an
equivalent grand canonical ensemble (see Appendix A). It was shown\cite
{Sitotaw Serota I}, however, that such a relationship holds only if the
particle-number fluctuation is large, which is the case when $T\gg \Delta $
also. A perturbative evaluation\cite{Oh Zyuzin Serota} using Imry's
formalism yields a persistent current $\propto T^{-1}$. The validity of this
result is obviously limited to temperatures $T\gtrsim \Delta $.

Evaluation of the orbital magnetic response for $T<\Delta $ is a challenging
problem. Unlike spin magnetism, which can be analyzed in terms of a
few-level problem\cite{Sitotaw Serota II}, all states of the Fermi sea
contribute, in principle, and there are large cancellations between the
diamagnetic and paramagnetic components. As a first step, however, one can
use a ''mixed approach'' wherein Imry's formalism is combined with the exact
form of the level-density correlation function. Such an approach should,
hopefully, reveal a trend in the persistent current as $T$ becomes smaller
than $\Delta $.

Two earlier attempts of this nature were undertaken in Refs.\cite{Altland et
al} and\cite{Serota Zyuzin}. While both correctly argued that the apparent
singularity for $T\rightarrow 0$ found in the perturbative approximation
disappears for GOE as the persistent current saturates to a finite value as $%
T$ crosses $\Delta $, they bore serious deficiencies.\ The former did not
find a closed form of the correlation function and considered only $T=0$
(while Imry's formalism, as was pointed out above, is rigorously applicable
when $T\gg \Delta $). The latter considered the entire range of temperatures
but the magnetic field dependence of the correlation function was crudely
mimicked by a substitution $\gamma \rightarrow \gamma +{{\tau }_{H}}^{-1}$
(as in perturbative expression (\ref{R_pert})) in the exact correlation
function obtained in the assumption of presence of level broadening $\gamma $
alone. Later, however, the ''mixed approach'' was fully implemented for spin
magnetism\cite{Sitotaw Serota I} where the magnetic field dependence of the
level density correlation function is rather simple.

It turns out, however, that the exact correlation function in the presence
of a time-reversal symmetry breaking field, such as magnetic field, had been
calculated over a decade earlier by Mehta and Pandey\cite{Mehta Pandey} for
both GOE\ and GSE using random matrix theory (RMT) (their result was
subsequently confirmed using the supersymmetric non-linear $\sigma $-model
in Ref.\cite{Altland Iida Efetov}). In this work we combine the exact
correlation function obtained in Refs.\cite{Mehta Pandey} and\cite{Altland
Iida Efetov} with Imry's formalism to find the magnetic field dependence of
the persistent current for finite temperatures.

\section{Level correlation function}

In units of $\left\langle \rho \right\rangle ^{2}$, the dependence of the
correlation function of the level density on the time reversal symmetry
breaking parameter $\lambda $ in transition from GOE to GUE (Gaussian
Unitary Ensemble) and GSE to GUE is given by\cite{Mehta Pandey}: 
\begin{eqnarray}
R_{orth}\left( x,\lambda \right) &=&\int_{1}^{\infty
}dk\int_{0}^{1}dk^{\prime }{\frac{k^{\prime }}{k}\sin \left( kx\right) \sin
\left( k^{\prime }x\right) e^{-2{\lambda }^{2}\pi ^{2}\left( k^{2}-k^{\prime
}{}^{2}\right) }}  \label{R_orth_exact} \\
R_{symp}\left( x,\lambda \right) &=&\frac{1}{4}\int_{1}^{\infty
}dk\int_{0}^{1}dk^{\prime }{\frac{k}{k^{\prime }}\sin \left( kx\right) \sin
\left( k^{\prime }x\right) e^{-2{\lambda }^{2}\pi ^{2}\left( k^{2}-k^{\prime
}{}^{2}\right) }}  \label{R_symp_exact}
\end{eqnarray}
respectively, where $x=\frac{\pi }{\Delta }|{\varepsilon }_{1}-{\varepsilon }%
_{2}|$. For a ring enclosing the flux $\phi $\cite{Altland Iida Efetov}, 
\begin{equation}
{\lambda }^{2}=\left( 4\pi \Delta {\tau }_{H}\right) ^{-1}  \label{lambda}
\end{equation}
where ${{\tau }_{H}}^{-1}$ is given by eq. (\ref{tau_H^-1}).

It follows from eqs. (\ref{R_orth_exact}) and (\ref{R_symp_exact}) that 
\begin{eqnarray}
R_{orth}\left( x,0\right) &=&-\frac{d}{dx}\left( \frac{\sin \left( x\right) 
}{x}\right) \left( \frac{\pi }{2}-%
\mathop{\rm Si}%
\left( x\right) \right)  \label{R_orth_exact_0} \\
R_{symp}\left( x,0\right) &=&\frac{1}{4}\left( \pi \delta \left( x\right) +%
\frac{d}{dx}\left( \frac{\sin \left( x\right) }{x}\right) 
\mathop{\rm Si}%
\left( x\right) \right)  \label{R_symp_exact_0}
\end{eqnarray}
The main difference between GOE and GSE is the $\delta $-function term in $%
R_{symp}\left( x,0\right) $. More precisely, it is a $\lambda \rightarrow 0$
limit of\cite{Kravtsov Zirnbauer}

\begin{equation}
\frac{\pi ^{\frac{5}{2}}x^{2}}{4\lambda ^{\frac{3}{2}}}\exp \left( -\frac{
\pi ^{2}x^{2}}{4\lambda }\right)  \label{delta_split_lambda}
\end{equation}
The latter will prove to be responsible for the singular temperature
dependence of the persistent current in GSE in the ''mixed approach.''

Consider now the full correlation function of the level density\cite{Mehta
Pandey},

\begin{eqnarray}
{\cal R}_{orth}\left( x,\lambda \right) &=&\pi \delta \left( x\right) -\frac{
\sin ^{2}\left( x\right) }{x^{2}}+R_{orth}\left( x,\lambda \right)
\label{Rcalig_orth_exact} \\
{\cal R}_{symp}\left( x,\lambda \right) &=&\frac{1}{4}\left( \pi \delta
\left( x\right) -\frac{\sin ^{2}\left( x\right) }{x^{2}}\right)
+R_{symp}\left( x,\lambda \right)  \label{Rcalig_symp_exact}
\end{eqnarray}
and notice that for either ensemble $R\left( x,\lambda \right) \rightarrow 0$
when $\lambda \rightarrow \infty $. It is clear then, using eqs. (\ref
{R_orth_exact_0}) and (\ref{R_symp_exact_0}), that GOE, GSE, and GUE\ all
exhibit a property of ''level rigidity,'' as manifested by

\begin{equation}
\int {\cal R}\left( x,0\right) dx=\int {\cal R}\left( x,\infty \right) dx=0
\label{level_ridgidity}
\end{equation}
(in fact, it is likely that $\int {\cal R}\left( x,\lambda \right) dx=0$ for 
{\it any} $\lambda $).

\section{Persistent currents}

As was explained in Refs.\cite{Imry} and\cite{Oh Zyuzin Serota}, the
dominant contribution to the mean orbital magnetic response of isolated
metals is given by the part of the magnetic free energy which differentiates
a canonical ensemble from an equivalent grand canonical ensemble (whose mean
particle number coincides with the fixed particle number of a canonical
ensemble). For $T\gg \Delta $, it is given by eq. (\ref{¶FH}) of Appendix A, 
\begin{eqnarray}
\delta F_{H} &=&\frac{1}{2\left\langle \rho \right\rangle }\partial _{H}\int
\int f\left( \varepsilon _{1},\mu \right) f\left( \varepsilon _{2},\mu
\right) R\left( \varepsilon _{1},\varepsilon _{2},{{\tau }_{H}}^{-1}\right)
d\varepsilon _{1}d\varepsilon _{2}  \nonumber \\
&=&-\frac{1}{2\left\langle \rho \right\rangle }\partial _{H}\int d\omega 
\frac{\omega }{\exp \left( \frac{\omega }{T}\right) -1}R\left( \omega ,{{\
\tau }_{H}}^{-1}\right)  \label{¶FH_main}
\end{eqnarray}
where integration on $\left( \varepsilon _{1}+\varepsilon _{2}\right) $ was
performed. Once $\delta F_{H}$ is known, the persistent current and the
magnetic moment are obtained, respectively via the following thermodynamic
identities 
\begin{equation}
I=-\frac{\partial \delta F_{H}}{\partial \phi }\text{, }M=-\frac{\partial
\delta F_{H}}{\partial H}  \label{I}
\end{equation}
We will use the expressions for the correlation function given in the
previous Section to evaluate eqs. (\ref{¶FH_main}) and (\ref{I}).

\subsection{Perturbative limit}

Here we address how the perturbation theory limit, whose leading term is
given by eq. (\ref{R_pert}), is recovered from eqs. (\ref{R_orth_exact}) and
(\ref{R_symp_exact}) (as was already mentioned, in what follows we will
neglect $\gamma $). This limit is important for the evaluation of the
orbital magnetic response when $T\gg \Delta $. The linear-response part, in
particular, is defined by the term $\propto {\tau }_{H}^{-1}$ in the free
energy. Notice, however, that the expansion of eq. (\ref{R_pert}) in orders
of ${\tau }_{H}^{-1}$ yields no linear contribution. The contribution linear
in ${\tau }_{H}^{-1}$ {\em in the level correlation function} comes, in
fact, from the next order term in expansion of eqs. (\ref{R_orth_exact}) and
(\ref{R_symp_exact}) and corresponds to the three-Cooperon diagrams, while
it is the two-Cooperon diagrams that yield eq. (\ref{R_pert}). On the other
hand, we will see that {\em in the expression for the free energy} such an
expansion of the correlation function is not well defined and a careful
analysis shows that the two-Cooperon diagrams give a leading contribution to
the linear response, while the contribution of the three-Cooperon diagrams
is smaller in the parameter $\Delta /T$.

Eq. (\ref{R_orth_exact}) can be written as 
\begin{eqnarray}
R_{orth}\left( x,\lambda \right) &\approx &%
\mathop{\rm Im}%
\left\{ \int_{1}{\frac{{e^{\left( ix-4{\lambda }^{2}\pi ^{2}\right) k{}}}}{k}%
}dk\right\} 
\mathop{\rm Im}%
\left\{ \int^{1}dk^{\prime }k^{\prime }{e^{\left( ix+4{\lambda }^{2}\pi
^{2}\right) k^{\prime }{}}}\right\}  \label{R_orth_pert_eval-1} \\
&=&\frac{1}{4}\left[ {e^{ix-4{\lambda }^{2}\pi ^{2}{}}}\left( \frac{{1}}{ix-4%
{\lambda }^{2}\pi ^{2}}+\frac{{1}}{\left( ix-4{\lambda }^{2}\pi ^{2}\right)
^{2}}\right) -c.c.\right] \left[ {e^{ix+4{\lambda }^{2}\pi ^{2}{}}}\left( 
\frac{{1}}{ix+4{\lambda }^{2}\pi ^{2}}-\frac{{1}}{\left( ix+4{\lambda }%
^{2}\pi ^{2}\right) ^{2}}\right) -c.c.\right]  \nonumber
\end{eqnarray}
where we used the fact that for the large values of $\left| ix-4{\lambda }%
^{2}\pi ^{2}\right| $ the main contribution will come from $k,k^{\prime
}\sim 1$, so that $k{}^{2}\approx 2k-1$, $k^{\prime }{}^{2}\approx
2k^{\prime }-1$ and used integration by parts. Omitting oscillating terms in
eq. (\ref{R_orth_pert_eval-1}), we find 
\begin{equation}
R_{orth}\left( x,\lambda \right) \approx \frac{1}{2}%
\mathop{\rm Re}%
\left\{ \frac{1}{\left( -ix+4{\lambda }^{2}\pi ^{2}\right) ^{2}}-\frac{2}{%
\left( -ix+4{\lambda }^{2}\pi ^{2}\right) ^{3}}\right\}
\label{R_orth_pert_eval-2}
\end{equation}
Recalling that $\left\langle \rho \right\rangle \Delta =s$, we finally
obtain 
\begin{equation}
R_{orth}\left( \omega ,{{\tau }_{H}}^{-1}\right) \approx \frac{s^{2}}{2\beta
\pi ^{2}}%
\mathop{\rm Re}%
\left\{ \frac{1}{\left( -i\omega +{{\tau }_{H}}^{-1}\right) ^{2}}-\frac{2}{%
\left( -i\omega +{{\tau }_{H}}^{-1}\right) ^{3}}\right\} \text{, }\beta =1
\label{R_orth_pert}
\end{equation}
A similar calculation yields 
\begin{equation}
R_{symp}\left( \omega ,{{\tau }_{H}}^{-1}\right) \approx \frac{s^{2}}{2\beta
\pi ^{2}}%
\mathop{\rm Re}%
\left\{ \frac{1}{\left( -i\omega +{{\tau }_{H}}^{-1}\right) ^{2}}+\frac{2}{%
\left( -i\omega +{{\tau }_{H}}^{-1}\right) ^{3}}\right\} \text{, }\beta =4
\label{R_symp_pert}
\end{equation}
The first term in eqs. (\ref{R_orth_pert}) and (\ref{R_symp_pert} coincides
with eq. (\ref{R_pert}) and the second term represents the three-Cooperon
diagrams.

Substituting eq. (\ref{R_pert}) in eq. (\ref{¶FH_main}), we find

\begin{equation}
\delta F_{H}=-\frac{1}{2\left\langle \rho \right\rangle }\partial _{H}\int
d\omega R_{pert}\left( \omega ,{{\tau }_{H}}^{-1}\right) \frac{\omega }{\exp
\left( \frac{\omega }{T}\right) -1}  \label{¶FH_pert}
\end{equation}
Expansion of $R_{pert}$ in terms of ${{\tau }_{H}}^{-1}$ would lead to
divergence of the integral at zero. Consequently, one needs to treat
carefully the second order pole at -${{i\tau }_{H}}^{-1}$. Alternatively,
one can change the order of $\int $ and $Re$ and the pole will be of the
first order. Using the latter procedure, one can close the contour in the
plane that does not contain the pole and evaluate the integral in terms of
the sum over Matsubara frequencies, $\omega _{m}=2\pi imT$. Since each
summand is $\propto m$ and ${{\tau }_{H}}^{-1}$ is finite at this stage of
the evaluation, the $m=0$ term yields zero and the expansion in orders of ${{%
\tau }_{H}}^{-1}$ is possible in the $m\geq 1$ Matsubara sum. As a result,
one finds the formulae derived in Ref.\cite{Oh Zyuzin Serota}. The second
term in eqs. (\ref{R_orth_pert} ) and (\ref{R_symp_pert}) can be treated
similarly and, as was mentioned above, gives a correction of the order of $%
\Delta /T$.

For completeness, a detailed perturbative evaluation of the persistent
current is given in Appendix B. Its main conclusion is that the persistent
current is given by 
\begin{eqnarray}
I_{pert}\left( \phi \right) &\approx &\frac{s^{2}}{2\beta \pi ^{2}}\frac{
8\varsigma \left( 2\right) eE_{c}}{T\left\langle \rho \right\rangle }\frac{
\phi }{\phi _{0}}\text{, }\phi \ll \phi _{c}  \label{I_pert_main-1} \\
I_{pert}\left( \phi \right) &\approx &\frac{s^{2}}{2\beta \pi ^{2}}\frac{e}{
2\pi \left\langle \rho \right\rangle }\frac{\phi _{0}}{\phi }\text{, }\phi
_{c}\ll \phi \ll \phi _{0}  \label{I_pert_main-2}
\end{eqnarray}
where 
\begin{equation}
\phi _{c}\sim \sqrt{\frac{T}{4\pi E_{c}}}\frac{\phi _{0}}{2}\ll \phi _{0}
\label{phi_c_main}
\end{equation}
is the magnetic flux scale for GOE\ to GUE\ transition. We emphasize the
universal character of the decay of the persistent current with the flux in
eq. (\ref{I_pert_main-2})\ first noticed in Ref.\cite{Oh Zyuzin Serota}.

\subsection{''Mixed approach''}

We now proceed to use the exact level correlation functions (\ref
{R_orth_exact}) and (\ref{R_symp_exact}) in eq. (\ref{¶FH_main}). While the
latter is valid for $T\gg \Delta $, in which case the perturbative
expression (\ref{R_pert}) for the correlation function is sufficient, we
hope that the results obtained in such an approach will be qualitatively
indicative of the behavior for $T<\Delta $.

\subsubsection{Orthogonal ensemble}

For the orthogonal ensemble eq. (\ref{I}) yields, with the help of eq. (\ref
{lambda}), the following expression for the persistent current 
\begin{eqnarray}
I_{orth}\left( \phi \right) &=&\frac{{\pi }^{3}}{2}\frac{\partial \tau
_{H}^{-1}}{\partial \phi }\frac{T^{2}\left\langle \rho \right\rangle }{%
\Delta }  \nonumber \\
&&\sum\limits_{m=1}^{\infty }m\int_{1}^{\infty }dk\int_{0}^{1}dk^{\prime }%
\frac{k^{\prime }}{k}\left( k^{2}-k^{\prime }{}^{2}\right) \left( e^{-2\pi
^{2}{m}\frac{T}{\Delta }\left( k-k^{\prime }\right) }-e^{-2\pi ^{2}{m}\frac{T%
}{\Delta }\left( k+k^{\prime }\right) }\right) e^{-\frac{1}{2}\pi \left(
\Delta \tau _{H}\right) ^{-1}\left( k^{2}-k^{\prime }{}^{2}\right) }
\label{I_orth_m}
\end{eqnarray}
where the integral over $\omega $ was reduced, by means of contour
integration, to summation over the Matsubara frequencies $\omega _{m}=2\pi
mT $, $m=1,2,3...$ . Performing summation over $m$, we obtain 
\begin{eqnarray}
I_{orth}\left( \phi \right) &=&\frac{{\pi }^{3}}{8}\frac{\partial \tau
_{H}^{-1}}{\partial \phi }\frac{T^{2}\left\langle \rho \right\rangle }{
\Delta }  \nonumber \\
&&\int_{1}^{\infty }dk\int_{0}^{1}dk^{\prime }\frac{k^{\prime }}{k}\left(
k^{2}-k^{\prime }{}^{2}\right) \left( \frac{1}{\sinh ^{2}\left( {\pi }^{2}%
\frac{T}{\Delta }\left( k-k^{\prime }\right) \right) }-\frac{1}{\sinh
^{2}\left( {\pi }^{2}\frac{T}{\Delta }\left( k+k^{\prime }\right) \right) }
\right) e^{-\frac{1}{2}\pi \left( \Delta \tau _{H}\right) ^{-1}\left(
k^{2}-k^{\prime }{}^{2}\right) }  \label{I_orth}
\end{eqnarray}
The integral in eq. (\ref{I_orth}) cannot be, in general, evaluated
analytically and we proceed to analyze some limiting cases which can be
explicitly handled.

\paragraph{Perturbative limit, $T\gg \Delta $}

While we have already discussed the perturbation theory limit, $T\gg \Delta $
, it is tutorial to derive the latter directly from eq. (\ref{I_orth}). To
this effect, omitting the exponentially small term with $\left( k+k^{\prime
}\right) $ and taking into account the fact that the main contribution comes
from $k,k^{\prime }\sim 1$, we find 
\begin{eqnarray}
I_{orth}\left( \phi \right) &\approx &\frac{{\pi }^{3}}{4}\frac{%
T^{2}\left\langle \rho \right\rangle }{\Delta }\frac{\partial \tau _{H}^{-1}%
}{\partial \phi }\int_{1}dk\int^{1}dk^{\prime }\frac{\left( k-k^{\prime
}{}\right) }{\sinh ^{2}\left( {\pi }^{2}\frac{T}{\Delta }\left( k-k^{\prime
}\right) \right) }e^{-\pi \left( \Delta \tau _{H}\right) ^{-1}\left(
k-k^{\prime }{}\right) }  \nonumber \\
&=&\frac{{\pi }^{3}}{4}\frac{T^{2}\left\langle \rho \right\rangle }{\Delta }%
\frac{\partial \tau _{H}^{-1}}{\partial \phi }\int_{0}^{\infty }du\frac{u}{
\sinh ^{2}\left( {\pi }^{2}\frac{T}{\Delta }u\right) }e^{-\pi \left( \Delta
\tau _{H}\right) ^{-1}u}\int_{-u{}}^{0}dv  \label{I_orth_pert_eval}
\end{eqnarray}
where we first shifted the variables of integration by $1$ (since the
integrand depends only on $k-k^{\prime }{}$) and then introduced the new
variables $u=\left( k-k^{\prime }\right) $ and $v=k^{\prime }$ and extended
the upper limit of integration to $\infty $ due to the exponentially
decaying integrand. Re-scaling the variable of integration, $x=\left( \pi
T/\Delta \right) u$, we finally arrive at eq. (\ref{I_pert_integral}) in
Appendix B.

\paragraph{Non-perturbative limit, $T\ll \Delta $}

Expanding the $\sinh $-functions in eq. (\ref{I_orth}), we find 
\begin{equation}
I_{orth}\left( \phi \right) \approx \frac{s}{2\pi }\frac{\partial \tau
_{H}^{-1}}{\partial \phi }\int_{1}^{\infty }dk\int_{0}^{1}dk^{\prime }\frac{%
k^{\prime 2}}{\left( k^{2}-k^{\prime }{}^{2}\right) }e^{-\frac{1}{2}\pi
\left( \Delta \tau _{H}\right) ^{-1}\left( k^{2}-k^{\prime }{}^{2}\right) }
\label{I_orth_T=0}
\end{equation}
This persistent current is the {\em limiting current in the zero-temperature
limit}. Notice, that the upper limit of the integral on $k$ has been
extended to $\infty $ (despite the small-parameter expansion of the
integrand), which introduces a correction of order $\max \left\{ \pi
^{2}T/\Delta ,\left( \frac{1}{2}\pi \tau _{H}^{-1}/\Delta \right)
^{1/2}\right\} $ to eq. (\ref{I_orth_T=0}). As in the perturbative case, eq.
(\ref{I_orth_T=0}) has the following limiting behaviors: 
\begin{eqnarray}
I_{orth}\left( \phi \right) &\approx &\frac{s}{2\pi }\frac{\partial \tau
_{H}^{-1}}{\partial \phi }=8eE_{c}\frac{\phi }{\phi _{0}}\text{, }\phi \ll
\phi _{c}  \label{I_orth-1} \\
I_{orth}\left( \phi \right) &\approx &\frac{s^{2}}{2\beta \pi ^{2}}\frac{e}{%
2\pi \left\langle \rho \right\rangle }\frac{\phi _{0}}{\phi }\text{, }\phi
_{c}\ll \phi \ll \phi _{0}  \label{I_orth-2}
\end{eqnarray}
where now $\phi _{c}$ is found from the condition $\frac{1}{2}\pi \tau
_{H}^{-1}/\Delta \sim 1$ as 
\begin{equation}
\phi _{c}\sim \sqrt{\frac{\Delta }{2\pi ^{3}E_{c}}}\frac{\phi _{0}}{2}\ll
\phi _{0}  \label{phi_c_orth}
\end{equation}
and we used the fact that 
\[
\int_{1}^{\infty }dk\int_{0}^{1}dk^{\prime }\frac{k^{\prime 2}}{\left(
k^{2}-k^{\prime }{}^{2}\right) }=\frac{1}{2} 
\]
and took explicitly $s=2$ in eq. (\ref{I_orth-1}). Notice that the universal
decay of the persistent current with the flux for $\phi _{c}\ll \phi \ll
\phi _{0}$ coincides with the perturbative result of eq. (\ref{I_pert_main-2}%
): this is obvious both on physical and mathematical grounds and comes from $%
k\sim k^{\prime }\sim 1$.

\subsubsection{Symplectic ensemble}

For GSE, we obtain, analogously to GOE, the following expression for the
persistent current 
\begin{eqnarray}
I_{symp}\left( \phi \right) &=&\frac{{\pi }^{3}}{8\beta }\frac{\partial \tau
_{H}^{-1}}{\partial \phi }\frac{T^{2}\left\langle \rho \right\rangle }{%
\Delta }  \nonumber \\
&&\int_{1}^{\infty }dk\int_{0}^{1}dk^{\prime }\frac{k}{k^{\prime }}\left(
k^{2}-k^{\prime }{}^{2}\right) \left( \frac{1}{\sinh ^{2}\left( {\pi }^{2}%
\frac{T}{\Delta }\left( k-k^{\prime }\right) \right) }-\frac{1}{\sinh
^{2}\left( {\pi }^{2}\frac{T}{\Delta }\left( k+k^{\prime }\right) \right) }
\right) e^{-\frac{1}{2}\pi \left( \Delta \tau _{H}\right) ^{-1}\left(
k^{2}-k^{\prime }{}^{2}\right) }  \label{I_symp}
\end{eqnarray}
As for GOE, we proceed to investigate the limiting cases of this expression.

\paragraph{Perturbative limit, $T\gg \Delta $}

This case is identical, up to coefficient $\beta ^{-1}$, to the orthogonal
ensemble since the reversed ratio $k/k^{\prime }$ relative to the latter is
inconsequential in view of the fact that the main contribution to the
integral comes from $k\sim k^{\prime }\sim 1$. As a result, we again find
that the expression for the persistent current given in Appendix B, with the
limiting behaviors described by eqs. (\ref{I_pert_main-1}) and (\ref
{I_pert_main-2}).

\paragraph{Non-perturbative limit, $T\ll \Delta $}

As for GOE, expanding the $\sinh $-functions in eq. (\ref{I_symp}), we find 
\begin{equation}
I_{symp}\left( \phi \right) \approx \frac{s}{2\beta \pi }\frac{\partial \tau
_{H}^{-1}}{\partial \phi }\int_{1}^{\infty }dk\int_{0}^{1}dk^{\prime }\frac{%
k^{2}}{\left( k^{2}-k^{\prime }{}^{2}\right) }e^{-\frac{1}{2}\pi \left(
\Delta \tau _{H}\right) ^{-1}\left( k^{2}-k^{\prime }{}^{2}\right) }
\label{I_symp_T=0}
\end{equation}
which is the {\em limiting current in the zero-temperature limit}. We first
address the zero-field limit. Towards this end, we notice that 
\begin{equation}
\int_{1}^{\infty }dk\int_{0}^{1}dk^{\prime }\frac{k^{2}}{\left(
k^{2}-k^{\prime }{}^{2}\right) }e^{-\alpha \left( k^{2}-k^{\prime
}{}^{2}\right) }\stackrel{\alpha \longrightarrow 0}{\longrightarrow }\frac{1%
}{2}\sqrt{\frac{\pi }{\alpha }}  \label{identity_alpha}
\end{equation}
which can be easily obtained by differentiation on $\alpha $ and subsequent
evaluation of the integral in terms of the Error Functions. Consequently, we
obtain 
\begin{equation}
I_{symp}\left( 0\right) \approx \frac{s}{2\beta \pi }\frac{\partial \tau
_{H}^{-1}}{\partial \phi }\frac{1}{2}\sqrt{\frac{2\Delta }{\tau _{H}^{-1}}}=%
\frac{se}{\beta \pi }\sqrt{2\Delta E_{c}}\text{, }T=0  \label{I_symp_0}
\end{equation}
The latter indicates the existence of the permanent current (and orbital
magnetic moment) for GSE, as first conjectured by Kravtsov and Zirnbauer\cite
{Kravtsov Zirnbauer}.

The permanent current in a zero flux implies a {\em Curie-type response} for
non-zero temperature. To obtain the latter, we need to be more careful when
expanding the $\sinh $-functions in eq. (\ref{I_symp}). Namely, we start by
expanding in ${\pi }^{2}\frac{T}{\Delta }k^{\prime }\ll 1$ which yields 
\begin{eqnarray}
I_{symp}\left( \phi \right) &\approx &\frac{{\pi }^{5}}{2\beta }\frac{%
\partial \tau _{H}^{-1}}{\partial \phi }\frac{T^{3}\left\langle \rho
\right\rangle }{\Delta ^{2}}\int_{1}^{\infty }dk\int_{0}^{1}dk^{\prime
}k\left( k^{2}-k^{\prime }{}^{2}\right) \frac{\cosh \left( {\pi }^{2}\frac{T%
}{\Delta }k\right) }{\sinh ^{3}\left( {\pi }^{2}\frac{T}{\Delta }k\right) }%
e^{-\frac{1}{2}\pi \left( \Delta \tau _{H}\right) ^{-1}\left(
k^{2}-k^{\prime }{}^{2}\right) }  \label{I_symp _T>0} \\
&\approx &\frac{{\pi }^{5}}{2\beta }\frac{\partial \tau _{H}^{-1}}{\partial
\phi }\frac{T^{3}\left\langle \rho \right\rangle }{\Delta ^{2}}%
\int_{1}^{\infty }dkk\frac{\cosh \left( {\pi }^{2}\frac{T}{\Delta }k\right) 
}{\sinh ^{3}\left( {\pi }^{2}\frac{T}{\Delta }k\right) }\int_{0}^{1}dk^{%
\prime }\left( k^{2}-k^{\prime }{}^{2}\right) \text{, }\phi \ll \frac{\phi
_{0}}{2}T\sqrt{\frac{2\pi }{\Delta E_{c}}}
\label{I_symp_T>0_linear_response}
\end{eqnarray}
where the latter condition on the flux gives the range of applicability of
the linear response for GSE in the limit $T\ll \Delta $. Rescaling the
variable of integration and omitting terms small in ratio $\Delta /T$, we
find 
\begin{equation}
I_{symp}\left( \phi \right) \approx \frac{{1}}{2{\pi }^{3}\beta }\frac{%
\partial \tau _{H}^{-1}}{\partial \phi }\frac{\Delta ^{2}\left\langle \rho
\right\rangle }{T}\int_{{0}}^{\infty }dyy^{3}\frac{\cosh \left( {y}\right) }{%
\sinh ^{3}\left( {y}\right) }=\frac{2se}{\beta }\frac{\Delta }{T}E_{c}\frac{%
\phi }{\phi _{0}}  \label{I_symp_linear_response}
\end{equation}
From the latter, $I_{symp}\left( \phi \right) \sim $ $\left( se/\beta
\right) \sqrt{2\pi \Delta E_{c}}$ at $\phi \sim \left( \phi _{0}/2\right) T%
\sqrt{2\pi /\left( \Delta E_{c}\right) }$ - in qualitative agreement with
eq. (\ref{I_symp_0}).

\section{Discussion}

The most interesting result is a possibility of a Curie-like orbital
response in GSE at temperatures smaller than the mean energy-level spacing.
In the ''mixed approach'' considered in this work, it appears as a result of
broadening by the magnetic flux of a $\delta $-function term in the level
density correlation function\cite{Kravtsov Zirnbauer}, in accordance with
eq. (\ref{delta_split_lambda}). While this broadening was attributed\cite
{Kravtsov Zirnbauer} to lifting of the Kramers degeneracy\cite{LL} by the
magnetic field, the level density description of the spectrum is, obviously,
insensitive to the differences between even- and odd-electron systems (this
equally applies to the use of eq. (\ref{¶FH_main}) which assumes large
variations of the particle number in the equivalent grand canonical
ensemble). On the other hand, the Kramers theorem distinguishes between
even- and odd-electron systems. In the odd case, the Curie orbital response
would be due to a permanent persistent current (moment) in a state
''unmatched'' by a time-reversed state of a degenerate doublet. The ''mixed
approach'' would then mimic the average behavior between the even- and
odd-electron systems. The situation is in complete analogy with spin
magnetism where the ''mixed approach'' is indicative of the Curie response
but one needs to study\cite{Sitotaw Serota I},\cite{Sitotaw Serota II}
few-level systems at $T\ll \Delta $ to correctly account for odd/even
differences. A similar analysis of the orbital response is complicated by
the need to consider large Fermi sea cancellations between its diamagnetic
and paramagnetic components. While odd/even difference is not important for
GOE, since the Fermi level states are not current-carrying, the few-level
analysis of the van-Vleck response must be carried out for $T\ll \Delta $.
We will address these issues in a future work.

\section{Acknowledgment}

We thank Bernard Goodman for very useful discussions.

\appendix 

\section{Response of the canonical ensemble}

In this Appendix we derive the relationship between the free energies of the
canonical ensmble and an equivalent grand canonical ensemble in the limit $%
T\gg \Delta $. Denoting $\partial _{H}$ the change of the appropriate
quantities with the magnetic field, we find

\begin{equation}
\partial _{H}F=\partial _{H}\Omega +N\partial _{H}\mu  \label{¶HF-1}
\end{equation}
where it is assumed that the chemical potential of a system may depend,
generally speaking, on the magnetic field. Using the idea that the magnetic
field dependence can be transferred from the energy level dependence to the
density of states \cite{Dingle},

\begin{equation}
\Omega =-T\int d\varepsilon \rho \left( \varepsilon ,H\right) \ln \left(
1+\exp \left( \frac{\mu -\varepsilon }{T}\right) \right)  \label{Omega}
\end{equation}
we find

\begin{equation}
\partial _{H}\Omega =-T\int d\varepsilon \partial _{H}\rho \left(
\varepsilon ,H\right) \ln \left( 1+\exp \left( \frac{\mu -\varepsilon }{T}
\right) \right) -\left( \partial _{H}\mu \right) N  \label{¶HOmega}
\end{equation}
where

\begin{equation}
N=\int d\varepsilon \rho \left( \varepsilon ,H\right) \left( 1+\exp \left( 
\frac{\varepsilon -\mu }{T}\right) \right) ^{-1}  \label{N-1}
\end{equation}
is the {\em fixed} number of particles in the canonical ensemble and the 
{\em mean} number of particles in the equivalent grand canonical ensemble.
Combining eqs. (\ref{¶HF-1}) - (\ref{N-1}), we obtain

\begin{equation}
\partial _{H}F=-T\int d\varepsilon \partial _{H}\rho \left( \varepsilon
,H\right) \ln \left( 1+\exp \left( \frac{\mu -\varepsilon }{T}\right) \right)
\label{¶HF-2}
\end{equation}
where the $\left( \partial _{H}\mu \right) N$ term, containing the
dependence of the chemical potential on the field, canceled out. Notice that
this is in agreement with the small corrections theorem wherein the
corrections to $F$ and $\Omega $ are the same, but in the latter we do not
vary $\mu $. The derivation that follows is insensitive to the dependence of 
$\mu $ on the magnetic field but only to the fact that it varies from system
to system\cite{Imry}.

Denoting $\left\langle \mu \right\rangle $ the{\em \ }average value of $\mu $
over all systems (disorder-average) in {\em zero field}, we proceed to
expand in powers of $\mu -\left\langle \mu \right\rangle $

\begin{eqnarray}
\partial _{H}F &\approx &-T\int d\varepsilon \partial _{H}\rho \left(
\varepsilon ,H\right) \ln \left( 1+\exp \left( \frac{\left\langle \mu
\right\rangle -\varepsilon }{T}\right) \right)  \nonumber \\
&&-\left( \mu -\left\langle \mu \right\rangle \right) \int d\varepsilon
\partial _{H}\left( \rho \left( \varepsilon ,H\right) -\left\langle \rho
\right\rangle \right) f\left( \varepsilon ,\left\langle \mu \right\rangle
\right)  \label{¶HFexpand}
\end{eqnarray}
where $f\left( \varepsilon ,\left\langle \mu \right\rangle \right) $ is the
Fermi distribution function and the first term has the meaning of $\partial
_{H}\Omega \left( \left\langle \mu \right\rangle \right) $. Taking into
account the following expression for the number of particles,

\begin{eqnarray}
N &=&\int d\varepsilon \left\langle \rho \right\rangle f\left( \varepsilon
,\left\langle \mu \right\rangle \right)  \nonumber \\
&=&\int d\varepsilon \rho \left( \varepsilon ,H\right) f\left( \varepsilon
,\mu \right)  \nonumber \\
&\approx &\int d\varepsilon \left\langle \rho \right\rangle f\left(
\varepsilon ,\left\langle \mu \right\rangle \right) +\int d\varepsilon
\left( \rho \left( \varepsilon ,H\right) -\left\langle \rho \right\rangle
\right) f\left( \varepsilon ,\left\langle \mu \right\rangle \right) -\left(
\mu -\left\langle \mu \right\rangle \right) \int d\varepsilon \left\langle
\rho \right\rangle f^{^{\prime }}\left( \varepsilon ,\left\langle \mu
\right\rangle \right)  \label{N-2}
\end{eqnarray}
and the fact that the last integral is $-\left\langle \rho \right\rangle ,$
we obtain, comparing the first and the last lines of the above equation,

\begin{equation}
\left( \mu -\left\langle \mu \right\rangle \right) \approx -\left\langle
\rho \right\rangle ^{-1}\int d\varepsilon \left( \rho \left( \varepsilon
,H\right) -\left\langle \rho \right\rangle \right) f\left( \varepsilon
,\left\langle \mu \right\rangle \right)  \label{¶mu}
\end{equation}
Substituting the latter in eq. (\ref{¶HFexpand}) and performing disorder
averaging, we find

\begin{eqnarray}
\delta F_{H} &\equiv &\partial _{H}F-\partial _{H}\Omega \left( \left\langle
\mu \right\rangle \right)  \nonumber \\
&\approx &\frac{1}{\left\langle \rho \right\rangle }\int \int d\varepsilon
_{1}d\varepsilon _{2}\left\langle \delta \rho \left( \varepsilon
_{1},H\right) \partial _{H}\delta \rho \left( \varepsilon _{2},H\right)
\right\rangle f\left( \varepsilon _{1},\left\langle \mu \right\rangle
\right) f\left( \varepsilon _{2},\left\langle \mu \right\rangle \right) 
\nonumber \\
&=&\frac{1}{2\left\langle \rho \right\rangle }\partial _{H}\int \int
d\varepsilon _{1}d\varepsilon _{2}\left\langle \delta \rho \left(
\varepsilon _{1},H\right) \delta \rho \left( \varepsilon _{2},H\right)
\right\rangle f\left( \varepsilon _{1},\left\langle \mu \right\rangle
\right) f\left( \varepsilon _{2},\left\langle \mu \right\rangle \right)
\label{¶FH} \\
&=&\frac{1}{2\left\langle \rho \right\rangle }\partial _{H}\left\langle
\delta N^{2}\right\rangle  \nonumber
\end{eqnarray}
where $\left\langle \delta N^{2}\right\rangle $ is the mean {\em particle
number fluctuation in the equivalent grand canonical ensemble}\cite{Landau
Lifshitz} described by $\Omega \left( \left\langle \mu \right\rangle \right) 
$.

The transformation in eq. (\ref{¶FH}) can be explained as follows. Notice
that $\left\langle \delta \rho \left( \varepsilon _{1},H_{1}\right) \partial
_{H_{2}}\delta \rho \left( \varepsilon _{2},H_{2}\right) \right\rangle $ can
be written as $\partial _{H_{2}}\left\langle \delta \rho \left( \varepsilon
_{1},H_{1}\right) \delta \rho \left( \varepsilon _{2},H_{2}\right)
\right\rangle $. However, the diffuson contribution to $\left\langle \delta
\rho \left( \varepsilon _{1},H_{1}\right) \delta \rho \left( \varepsilon
_{2},H_{2}\right) \right\rangle $ depends on $\left( H_{1}-H_{2}\right) ^{2}$
and the Cooperon contribution on $\left( H_{1}+H_{2}\right) ^{2}$(see, for
instance, Ref.\cite{Oh Zyuzin Serota}). Expanding in $\left( H_{1}\pm
H_{2}\right) ^{2}$ and taking the limit $H_{1}\rightarrow H_{2}=H$, we
arrive at the desired result.

All of the above results can be also obtained in a purely thermodynamical
framework using the small corrections theorem\cite{Landau Lifshitz}. Indeed,
the latter states that small corrections to all thermodynamic potentials are
the same provided that the appropriate thermodynamic variables are kept
constant. Therefore, we write,

\begin{equation}
\delta _{H}F=\delta _{H}\Omega \left( \mu \left( H\right) ,H\right)
\label{deltaHF-1}
\end{equation}
an no variation of $\mu \left( H\right) $ is done in the r.h.s. This is in
correspondence with eq. (\ref{¶FH}) where only the level density is varied
with the field. Expanding around $\left\langle \mu \right\rangle $, we find

\begin{equation}
\delta _{H}F\approx \delta _{H}\Omega \left( \left\langle \mu \right\rangle
,H\right) +\delta _{H}\left( \frac{\partial \Omega }{\partial \mu }\left(
\left\langle \mu \right\rangle ,H\right) \right) \left( \mu -\left\langle
\mu \right\rangle \right)  \label{deltaHF-2}
\end{equation}
which, again, is in complete correspondence with the expansion in eq. (\ref
{¶HFexpand}). Noticing that

\begin{equation}
N\left( \mu \left( H\right) ,H\right) =N  \label{N_constraint}
\end{equation}
we reproduce eq. (\ref{¶FH}) (neglecting the difference between the
thermodynamic and single-particle densities of states, $\partial N/\partial
\mu $ and $\left\langle \rho \right\rangle $ respectively, and after some
manipulations analogous to those in eq. (\ref{N-2})) as follows: 
\begin{eqnarray}
\delta F_{H} &\equiv &\delta _{H}F-\delta _{H}\Omega \left( \left\langle \mu
\right\rangle ,H\right)  \nonumber \\
&\approx &-\delta _{H}\left( N\left( \left\langle \mu \right\rangle
,H\right) -N\left( \mu \left( H\right) ,H\right) \right) \left( \mu
-\left\langle \mu \right\rangle \right)  \nonumber \\
&\approx &\frac{\partial N\left( \left\langle \mu \right\rangle ,H\right) }{%
\partial \mu }\left( \delta _{H}\left( \mu -\left\langle \mu \right\rangle
\right) \right) \left( \mu -\left\langle \mu \right\rangle \right)
\label{¶FH-2} \\
&\approx &\frac{\left\langle \rho \right\rangle }{2}\delta _{H}\left( \mu
-\left\langle \mu \right\rangle \right) ^{2}\approx \frac{1}{2\left\langle
\rho \right\rangle }\delta _{H}\left( N-\left\langle N\right\rangle \right)
^{2}  \nonumber
\end{eqnarray}

\section{Persistent current in the perturbative approximation}

Using eqs. (\ref{R_pert}), (\ref{¶FH_pert}) and (\ref{I}) we find, upon
reducing the integral to the Matsubara sum, 
\begin{equation}
I_{pert}\left( \phi \right) \ \ \approx -\frac{s^{2}}{2\beta \pi ^{2}}\frac{%
T^{2}}{2\left\langle \rho \right\rangle }\frac{\partial }{\partial \phi }%
\sum_{m=1}^{\infty }\frac{m}{\left[ mT+\left( 2\pi \tau _{H}\right)
^{-1}\right] ^{2}}  \label{I_pert_m}
\end{equation}
With the help of the identity $u^{-2}=\int_{0}^{\infty }t\exp \left(
-ut\right) dt$ the latter can be re-written as 
\begin{eqnarray}
I_{pert}\left( \phi \right) \  &=&-\frac{s^{2}}{2\beta \pi ^{2}}\frac{T^{2}}{%
2\left\langle \rho \right\rangle }\frac{\partial }{\partial \phi }\int_{0}^{%
\infty }\left( \sum_{m=1}^{\infty }me^{-mTt}\right) t\exp \left( -\frac{\tau
_{H}^{-1}}{2\pi }t\right) dt  \nonumber \\
&=&\frac{s^{2}}{2\beta \pi ^{2}}\frac{1}{2T\left\langle \rho \right\rangle }%
\frac{\partial \tau _{H}^{-1}}{\partial \phi }\int_{0}^{\infty }\left( \frac{%
\pi x}{\sinh \left( \pi x\right) }\right) ^{2}\exp \left( -\frac{\tau
_{H}^{-1}}{T}x\right) dx  \label{I_pert_integral}
\end{eqnarray}
which coincides with the respective equation\footnote{%
Notice that, due to a misprint, the factor $T^{-1}$ is missing in eq. (18)
of Ref.\cite{Oh Zyuzin Serota}.} in Ref.\cite{Oh Zyuzin Serota}. From eq. (%
\ref{I_pert_integral}) it follows that \cite{Oh Zyuzin Serota} 
\begin{eqnarray}
I_{pert}\left( \phi \right)  &\approx &\frac{s^{2}}{2\beta \pi ^{2}}\frac{1}{%
2\pi T\left\langle \rho \right\rangle }\frac{\partial \tau _{H}^{-1}}{%
\partial \phi }\int_{0}^{\infty }\left( \frac{x}{\sinh x}\right) ^{2}dx=%
\frac{s^{2}}{2\beta \pi ^{2}}\frac{8\varsigma \left( 2\right) eE_{c}}{%
T\left\langle \rho \right\rangle }\frac{\phi }{\phi _{0}}\text{, }\phi \ll
\phi _{c}  \label{I_pert-1} \\
I_{pert}\left( \phi \right)  &\approx &\frac{s^{2}}{2\beta \pi ^{2}}\frac{1}{%
2T\left\langle \rho \right\rangle }\frac{\partial \tau _{H}^{-1}}{\partial %
\phi }\int_{0}^{\infty }\exp \left( -\frac{\tau _{H}^{-1}}{T}x\right) dx=%
\frac{s^{2}}{2\beta \pi ^{2}}\frac{e}{2\pi \left\langle \rho \right\rangle }%
\frac{\phi _{0}}{\phi }\text{, }\phi _{c}\ll \phi \ll \phi _{0}
\label{I_pert-2}
\end{eqnarray}
where $\varsigma \left( 2\right) =\pi ^{2}/6$ and $\phi _{c}$ is found from
the condition $\tau _{H}^{-1}\sim T$ and is given by \cite{Oh Zyuzin Serota} 
\begin{equation}
\phi _{c}\sim \sqrt{\frac{T}{4\pi ^{2}E_{c}}}\frac{\phi _{0}}{2}\ll \phi _{0}
\label{phi_c}
\end{equation}

For completeness, we also derive the result accounting for the Aharonov-Bohm
periodicity in narrow rings. This is easily accomplished by using the
following expression for the perturbative level-density correlation function 
\cite{AS} 
\begin{equation}
R_{pert}\left( \omega ,{{\tau }_{H}}^{-1}\right) =\frac{s^{2}}{2\beta {\pi }%
^{2}}\sum_{n=-\infty }^{\infty }Re\frac{1}{{\left( -i\omega +4\pi %
^{2}E_{c}\left( n+\frac{2\phi }{\phi _{0}}\right) ^{2}\right) }^{2}}
\label{R_pert_AB}
\end{equation}
As a result, we find the following expression for the persistent current 
\begin{equation}
I_{pert}\left( \phi \right) \ \ \approx -\frac{s^{2}}{2\beta \pi ^{2}}\frac{%
T^{2}}{2\left\langle \rho \right\rangle }\frac{\partial }{\partial \phi }%
\sum_{n=-\infty }^{\infty }\sum_{m=1}^{\infty }\frac{m}{\left[ mT+2\pi %
E\left( n+\frac{2\phi }{\phi _{0}}\right) ^{2}\right] ^{2}}
\label{I_pert_AB_eval1}
\end{equation}
which can be evaluated using and identity 
\begin{equation}
\sum_{n=-\infty }^{\infty }\frac{1}{\left[ b^{2}+\left( n+a\right)
^{2}\right] }=\frac{\pi }{b}\frac{\sinh \left( 2\pi b\right) }{\cosh \left( 2%
\pi b\right) -\cos \left( 2\pi a\right) }  \label{identity}
\end{equation}
(notice that for small $a$ and $b$ the sum is dominated by the $n=0$ term -
in agreement with the procedure adopted in the bulk of the article). From
eqs. (\ref{I_pert_AB_eval1})\ and (\ref{identity}), we find\ 
\begin{equation}
I_{pert}\left( \phi \right) \approx \frac{s^{2}}{2\beta \pi ^{2}}\frac{\pi %
T^{2}}{2\left\langle \rho \right\rangle E_{c}}\frac{\partial }{\partial T}%
\frac{\partial }{\partial \phi }\sum_{m=1}^{\infty }\frac{1}{\sqrt{\frac{2%
\pi T}{E_{c}}m}}\frac{\sinh \left( \sqrt{\frac{2\pi T}{E_{c}}m}\right) }{%
\cosh \left( \sqrt{\frac{2\pi T}{E_{c}}m}\right) -\cos \left( \frac{4\pi %
\phi }{\phi _{0}}\right) }  \label{I_pert_AB_eval2}
\end{equation}
In Ref.\cite{Oh Zyuzin Serota}, the sum appearing in this equation\ was
evaluated using the Euler-Maclaurin method. However, the latter does not
work well for small values of the argument. Consequently, here we choose a
different approach. Namely, we use the fact that $2\pi T\ll E_{c}$ and
expand the argument for $m\ll E_{c}/2\pi T$ and extend the sum to infinity
thus neglecting the correction of the order $2\pi T/E_{c}$ 
\begin{eqnarray}
I_{pert}\left( \phi \right)  &\approx &\frac{s^{2}}{2\beta \pi ^{2}}\frac{%
2eE_{c}}{\pi T\left\langle \rho \right\rangle }\sin \left( 4\pi \phi /\phi
_{0}\right) [\psi ^{\left( 1\right) }\left( z\right) +\frac{1}{2}z\psi
^{\left( 2\right) }\left( z\right) ]  \label{I_pert_AB} \\
z &=&\frac{E_{c}}{\pi T}\left[ 1-\cos \left( 4\pi \phi /\phi _{0}\right)
\right]   \label{z}
\end{eqnarray}
where $\psi ^{\left( n\right) }\left( z\right) $ is the polygamma function.
Using $\psi ^{\left( 1\right) }\left( z\right) +\frac{1}{2}z\psi ^{\left(
2\right) }\left( z\right) \stackrel{z\longrightarrow 0}{\longrightarrow }%
\varsigma \left( 2\right) $ and $\psi ^{\left( 1\right) }\left( z\right) +%
\frac{1}{2}z\psi ^{\left( 2\right) }\left( z\right) \stackrel{%
z\longrightarrow \infty }{\longrightarrow }\left( 2z\right) ^{-1}$, we
recover - in the limit $\phi \ll \phi _{0}$ - eqs. (\ref{I_pert-1}) and (\ref
{I_pert-2})\ from eq. (\ref{I_pert_AB}).

\end{document}